\documentclass[12pt]{article}
\usepackage{epsfig,float,wrapfig}
\usepackage{color}
\usepackage{fancyhdr}
\usepackage{amsmath}
\usepackage{amssymb} 

\allowdisplaybreaks[4]

\newcommand{\be}{\begin{equation}}
\newcommand{\ba}{\begin{eqnarray}}
\newcommand{\ea}{\end{eqnarray}}

\def\a{\alpha}
\def\b{\beta}

\def\d{\delta}
\def\e{\epsilon}
\def\ve{\varepsilon}
\def\f{\phi}

\def\k{\kappa}
\def\l{\lambda}
\def\m{\mu}
\def\n{\nu}

\def\p{\pi}
\def\q{\theta}
\def\r{\rho}
\def\s{\sigma}

\def\OO{\Omega}
\def\P{\Pi}

\def\S{\Sigma}

\def\cf{{\cal F}}

\def\cs{{\cal S}}

\newcommand{\ti}{\tilde}

\newcommand{\pa}{\partial}

\newcommand{\ptt}{\Tilde{\Tilde{p}}}

\renewcommand{\title}[1]{\null\vspace{25mm}\noindent{\Large{\bf #1}}\vspace{10mm}}
\newcommand{\authors}[1]{\noindent{\large #1}\vspace{20mm}}
\newcommand{\address}[1]{{\center{\noindent #1\vspace{0mm}}}}
\renewcommand{\abstract}[1]{\vspace{17mm}
\noindent{\small{\em Abstract.} #1}\vspace{2mm}}

\DeclareMathAlphabet\mathbb  {U}{msb}{m}{n}
\DeclareFontFamily{U}{msb}{} \DeclareFontShape{U}{msb}{m}{n}{
  <5> <6> <7> <8> <9> gen * msbm
  <10> <10.95> <12> <14.4> <17.28> <20.74> <24.88> msbm10
  }{}

\makeatletter
\def\section{\@startsection{section}{1}{\z@}{-3.25ex plus -1ex minus
             -.2ex}{1.5ex plus .2ex}{\normalfont\bfseries}}
\def\subsection{\@startsection{subsection}{1}{\z@}{-3.25ex plus -1ex
                minus -.2ex}{1.5ex plus .2ex}{\normalfont\itshape}}

\renewenvironment{thebibliography}[1]
         {\section*{References}\frenchspacing\small
          \begin{list}{[\arabic{enumi}]}
         {\usecounter{enumi}\parsep=2pt\topsep 0pt
         \settowidth{\labelwidth}{[#1]}
         \leftmargin=\labelwidth\advance\leftmargin\labelsep
         \rightmargin=0pt\itemsep=0pt\sloppy}}{\end{list}}

\makeatother
\begin{document}   \setcounter{table}{0}
 
\begin{titlepage}
\begin{center}
\hspace*{\fill}{{\normalsize \begin{tabular}{l}
                              {\sf hep-th/0102044}\\
                              {\sf REF. TUW 01-01}\\
                              {\sf REF. UWThPh-2001-7}
                              \end{tabular}   }}

\title{\vspace{5mm} Perturbative Analysis of the Seiberg-Witten Map}

\authors {  \Large{A.A. Bichl$^{1}$, J. M. Grimstrup$^{2}$, L. Popp$^{3}$, M. Schweda$^{4}$,\\[1mm] R. Wulkenhaar$^{5}$ }}    \vspace{-20mm}

\vspace{10mm}
       
\address{$^{1,2,3,4}$  Institut f\"ur Theoretische Physik, Technische Universit\"at Wien\\
      Wiedner Hauptstra\ss e 8-10, A-1040 Wien, Austria}
\address{$^{5}$  Institut f\"ur Theoretische Physik, Universit\"at Wien\\Boltzmanngasse 5, A-1090 Wien, Austria   }

\footnotetext[1]{bichl@hep.itp.tuwien.ac.at, work supported in part by ``Fonds zur F\"orderung der Wissenschaftlichen Forschung'' (FWF) under contract P14639-PHY.}

\footnotetext[2]{jesper@hep.itp.tuwien.ac.at, work supported by The Danish Research Agency.}

\footnotetext[3]{popp@hep.itp.tuwien.ac.at, work supported in part by ``Fonds zur F\"orderung der Wissenschaftlichen Forschung'' (FWF) under contract P13125-PHY.}
\footnotetext[4]{mschweda@tph.tuwien.ac.at.}
\footnotetext[5]{raimar@doppler.thp.univie.ac.at, Marie-Curie Fellow.}   
\end{center}

\thispagestyle{empty}
\begin{center}
\begin{minipage}{12cm}

\vspace{10mm}

{\it Abstract.} We investigate the quantization of the $\q$-expanded noncommutative $U(1)$ Yang-Mills action, obtained via the Seiberg-Witten map. As expected we find non-renormalizable terms. The one-loop propagator corrections are gauge independent, and lead us to a unique extention of the noncommutative classical action. We interpret our results as a requirement that also the trace in noncommutative field theory should be deformed.

\end{minipage}\end{center}
\end{titlepage}

\section{Noncommutative Yang-Mills Theory and the Seiberg-Witten Map}

The Seiberg-Witten map was first discovered in the context of string theory, where it emerged from a 2D-$\s$-model regularized in different ways \cite{Seiberg:1999vs}. It was argued by Seiberg and Witten that the ordinary gauge theory should be gauge-equivalent to a noncommutative Yang-Mills (NCYM) field theory, which, in a certain limit, acts as an effective theory of open strings. Furthermore, they showed that the Seiberg-Witten map could be interpreted as an infinitesimal shift in the noncommutative parameter $\q$, and thus as an expansion of the noncommutative gauge field in $\q$.   

 Whereas in open string theory the (noncommutative) gauge fields are taken to transform in a certain matrix representation of a U(N) gauge group, the aim of a second approach to the subject \cite{Madore:2000en,Jurco:2000ja} was to realize a general, non-Abelian gauge group, preferably SU(N). Using covariant coordinates the NCYM theory emerges as the gauge theory of a certain noncommutative algebra \cite{Madore:2000en}. However, in this scenario, due to the choice of a general, non-Abelian gauge group, one is forced to consider enveloping algebra-valued fields, which leads to infinitely many degrees of freedom \cite{Jurco:2000ja}. The solution to this problem was shown to be the Seiberg-Witten map, which in this context appears as an expansion of the noncommutative gauge field in both $\q$ and the generators of the gauge group. Application of the Seiberg-Witten map yields a theory with finitely many degrees of freedom. However, since the Seiberg-Witten map is infinitely non-linear, the resulting theory has infinitely many interactions at arbitrary high orders in the gauge field. Furthermore, since the noncommutative parameter $\q$, which has dimension $-2$, appears as a coupling constant, the model is non-renormalizable in the traditional sense. In the following we will refer to this model as the $\q$-expanded NCYM.

 The aim of this paper is to study the quantization of the $\q$-expanded NCYM. We choose to consider the case of an Abelian, \emph{i.e.} U(1), gauge group: noncommutative Maxwell theory. 

 The question of quantization of apparently non-renormalizable theories has been addressed in the literature, see \emph{e.g.}\ \cite{Gomis:1996jp} and citations therein. As a starting point, one could speculate if a power-counting non-renormalizable theory involving infinitely many interactions at arbitrary order in the field, as it is the case in the $\q$-expanded NCYM theory, could indeed be renormalizable in the sense that all divergent graphs may be absorbed in the classical action. However, we find that this is not the case for the $\q$-expanded NCYM. The self-energy produces terms which cannot be renormalized, thus forcing us to add extra, gauge invariant, terms quadratic in $\q$ to the classical action of NCYM theory, yielding an extended NCYM theory. We regard this extention as the lowest order of an infinite deformation series of the scalar product. Furthermore, a consequence of the extended classical action is that propagation of light is altered. One may speculate whether this could lead to observable effects in \emph{e.g.}\ cosmology.

 One may object that an expansion in $\q$ is not adequate for the following two reasons. First of all, taking all orders of $\q$ into account, it was shown, in the context of string theory, that $\q$ serves as a regulator for non-planar graphs \cite{Filk:1996dm} rendering otherwise UV-divergent graphs finite. The resulting radiative correction, however, is divergent for $\q\rightarrow 0$, thus suggesting that the effective action is not analytical in $\q$ \cite{Minwalla:1999px}. Secondly, one could argue that renormalizability dictates one to take all orders of $\q$ into account. Whereas \emph{e.g.}\ the noncommutative $\f^{4}$-theory expanded to $n$'th order in $\q$ is obviously (perturbatively in the coupling constant) non-renormalizable, the theory is two-loop renormalizable \cite{Aref'eva:2000sn} when all orders of $\q$ are taken into account. However, if one insists on treating a general gauge group, the expansion in $\q$ is the only known method of obtaining a quantizable action. In fact one may ask the question of how a noncommutative (gauge) theories should be correctly quantized.

 The paper is organized as follows. In section 2 we give the classical action expanded to first order in $\q$. The gauge fixing is performed in section 3, where we argue that two fundamentally different ways of introducing ghosts to the theory, via a linear and a non-linear gauge, may be applied. In section 4 we give the relevant Feynman rules and calculate the self-energy to second order in $\q$. The extended NCYM theory is given in section 5, and in section 6 we present our summary and discussion.

\section{$\q$-expanded NCYM}

We consider the coordinates of a (flat) Minkowski space as self-adjoint operators on a Hilbert space with the following algebra
\be
\left[x_{\m},x_{\n}\right] = {\rm{i}}  \q_{\m\n},\label{com}
\end{equation}
where $\q_{\m\n}$ is real and antisymmetric. A field theory in this context is equivalent to a field theory on a usual (commutative) flat manifold with the product substituted by the non-local $\star$-product\footnote{We use the following Fourier conventions: $f(x)=\int \frac{d^{4}p}{(2\p)^{4}} e^{-{\rm{i}}p_{\m}x^{\m}}\ti{f}(p),\; \ti{f}(p)=\int d^{4}x e^{{\rm{i}}p_{\m}x^{\m}}f(x)$. }
\be
(f\star g)(x)=\int \frac{d^{4}k }{(2\p)^{4}}\int \frac{d^{4}p}{(2\p)^{4}} \,e^{-{\rm{i}}(k_{\m}+p_{\m})x^{\m}}e^{-\frac{{\rm{i}}}{2}\q^{\m\n}k_{\m}p_{\n}}\ti{f}(k)\ti{g}(p),\label{ny-1}
\end{equation}
where $f$ and $g$ are functions on the manifold. A $U(1)$ gauge field $\hat{A}_{\m}=\hat{A}_{\m}^{*}$ (Hermitian) gives rise to the noncommutative Yang-Mills action\footnote{There could be a coupling constant added, however, in the absence of $\q$-independent interactions this coupling constant is not renormalized and may be absorbed in a reparametrization.}
\be
\hat{\S}_{cl}=-\frac{1}{4}\int d^{4}x\, \hat{F}_{\m\n}\star \hat{F}^{\m\n}=-\frac{1}{4}\int d^{4}x\, \hat{F}_{\m\n}\hat{F}^{\m\n},\label{action}
\end{equation}
with
\be
\hat{F}_{\m\n}=\pa_{\m}\hat{A}_{\n}-\pa_{\n}\hat{A}_{\m}-{\rm{i}}\hat{A}_{\m}\star\hat{A}_{\n}+{\rm{i}}\hat{A}_{\n}\star\hat{A}_{\m}.
\end{equation}
The action (\ref{action}) is invariant under the noncommutative gauge transformation
\be
\hat{\d}_{\hat{\l}}\hat{A}_{\m}=\pa_{\m}\hat{\l}-{\rm{i}}\hat{A}_{\m}\star\hat{\l}+ {\rm{i}}\hat{\l}\star\hat{A}_{\m} \equiv \hat{D}_{\m}\hat{\l}\label{gaugeI},
\end{equation} 
with infinitesimal $\hat{\l}=\hat{\l}^{*}$. It was shown by Seiberg and Witten \cite{Seiberg:1999vs} that an expansion in $\q$ leads to a map between the noncommutative gauge field $\hat{A}_{\m}$ and the commutative gauge field $A_{\m}$ as well as their respective gauge parameters $\hat{\l}$ and $\l$, known as the Seiberg-Witten map:    
\ba
\hat{A}_{\m}\left(A\right)&=& A_{\m}- \frac{1}{2}\q^{\r\s} A_{\r}\left(\pa_{\s}A_{\m}+F_{\s\m}\right) + O(\q^{2}),\label{sw1}\\ 
\hat{\l}\left(\l,A\right)&=&\l-\frac{1}{2}\q^{\r\s}A_{\r}\pa_{\s}\l + O(\q^{2}),\label{sw2}
\ea
where the Abelian field strength is given by
\be
F_{\m\n}=\pa_{\m}A_{\n}-\pa_{\n}A_{\m}.
\end{equation}
Insertion of (\ref{sw1}) into (\ref{action}) leads to the action
\be
\S_{cl}=\int d^4x \left(-\frac{1}{4}F_{\m\n}F^{\m\n}+\frac{1}{8}\theta^{\a\b}F_{\a\b}F_{\m\n}F^{\m\n}-\frac{1}{2}\theta^{\a\b}F_{\m\a}F_{\n\b}F^{\m\n}    \right)+ O(\q^{2}),\label{actionII}
\end{equation}
which is invariant under the usual Abelian gauge transformations
\be
\d_{\l}A_{\m}=\pa_{\m}\l \label{gaugeII}.
\end{equation}
The action (\ref{actionII}) has in its full form, involving all orders of $\q$, infinitely many interactions at infinitely high order in the gauge field. Furthermore, since $\q$ has dimension $-2$, the theory is power-counting non-renormalizable in the traditional sense.

\section{Gauge Fixing}

In order to quantize a gauge theory within the BRST-scheme, the gauge-symmetry is replaced by the nilpotent BRST-symmetry \cite{Becchi:1974xu,Piguet:1995er}. However, above we have two gauge symmetries: $\hat{\d}_{\hat{\l}}$ and $\d_{\l}$ corresponding to the actions (\ref{action}) and (\ref{actionII}), respectively. Thus, there appear to be at least two fundamentally different ways of introducing ghosts into the theory, before and after performing the Seiberg-Witten map.

 Let us first consider the gauge-transformation (\ref{gaugeII}) as the ``fundamental'' one and introduce ghosts into the action (\ref{actionII}). We write
\be
sA_{\m} = \pa_{\m}c,\,\,\,\,\,\,
sc =0 ,\label{BRST1}
\end{equation}
where $s$ is the BRST-operator and $c$ the anti-commuting Faddeev-Popov ghost field. Within the quantization procedure a BRST-invariant gauge-fixing may be introduced in the following manner
\be
\S^{(i)}_{gf} =  \int d^{4}x \;\left[s\left( \bar{c} \pa_{\m} A^{\m} \right)+\frac{\a}{2}B^{2}\right],
\end{equation}
with
\be
s\bar{c} = B ,\,\,\,\,\, sB =0.\label{BRST2}
\end{equation}
Here $\bar{c}$ is the anti-ghost field and $B$ the Nakanishi-Lautrup (multiplier) field. 
The total action is now
\be
\S^{(i)}_{tot} = \S_{cl} + \S^{(i)}_{gf}.\label{ACTION-gf}
\end{equation}
In the following we will refer to this choice of gauge-fixing as {\it the linear gauge}.\\

 Let us now consider the second option of introducing ghosts in the theory. We treat the gauge transformation (\ref{gaugeI}) as the source of ghosts and thereby adding a gauge-fixing term to the action (\ref{action}). We write
\be
\hat{s}\hat{A}_{\m} = \hat{D}_{\m}\hat{c},\,\,\,\,\,\,
\hat{s}\hat{c} ={\rm{i}}\hat{c}\star\hat{c},
\end{equation}
where $\hat{s}$ is the BRST-operator emerging from the gauge-symmetry (\ref{gaugeI}) and $\hat{c}$ the corresponding ghost field. The gauge-fixing term reads
\be
\hat{\S}_{gf} =  \int d^{4}x\left[ \hat{s}\left( \hat{\bar{c}}\star \pa^{\m} \hat{A}_{\m} \right)+\frac{\a}{2}\hat{B}\star\hat{B}\right],
\end{equation}
with
\be
\hat{s}\hat{\bar{c}} = \hat{B} ,\,\,\,\,\, \hat{s}\hat{B} =0.
\end{equation}
Here $\hat{\bar{c}}$ and $\hat{B}$ are the anti-ghost and multiplier field. 
The total action is now
\be
\hat{\S}_{tot} = \hat{\S}_{cl} + \hat{\S}_{gf}.\label{ACTION-GFII}
\end{equation}
In order to apply the Seiberg-Witten map to (\ref{ACTION-GFII}) we need the Seiberg-Witten map of the ghost and multiplier field. These are easily found by substituting $\l$ with $c$ and $\hat{\l}$ with $\hat{c}$ in (\ref{sw2}). Notice that only the gauge field and the ghost have an expansion in $\q$:
\ba
\hat{c}\left(c\right)&=&c-\frac{1}{2}\q^{\n\m}A_{\n}\pa_{\m}c + O(\q^{2}),\label{sw3}\\ 
\hat{\bar{c}}&=&\bar{c},\label{sw4}\\
\hat{B}&=&B,\label{sw5}
\ea
where $c$, $\bar{c}$ and $B$ are the ordinary ghost, anti-ghost and multiplier field, respectively. Inserting (\ref{sw1}) and (\ref{sw3})--(\ref{sw5}) into (\ref{ACTION-GFII}) one finds, to first order in $\q$, the action
\be
\S^{(ii)}=  \S_{cl} +  \S^{(ii)}_{gf} ,         \label{ACTION-gfII}
\end{equation}
with
\ba
\S^{(ii)}_{gf}&=& \int d^{4}x \Big[ B\pa^{\m}A_{\m}-\bar{c}\pa_{\m}\pa^{\m}c \nonumber\\&&
-\q^{\a\b}\Big(\pa^{\m}\bar{c}\pa_{\a}c\pa_{\b}A_{\m}
-\frac{1}{2} \pa^{\m}\pa_{\m}\bar{c}A_{\a}\pa_{\b}c
-\frac{1}{2}  \pa^{\m}BA_{\a}\left(\pa_{\b}A_{\m}+F_{\b\m} \right)\Big)                   \Big]      , 
\ea
which is invariant under the BRST-transformations (\ref{BRST1}) and (\ref{BRST2}). Notice that (\ref{ACTION-gfII}) represents a nonlinear gauge.
In the following we will refer to this choice of gauge-fixing as {\it the nonlinear gauge}.\\

 Both gauge-fixed actions (\ref{ACTION-gf}) and (\ref{ACTION-gfII}) are invariant under Abelian BRST-transformations and satisfy the Slavnov-Taylor identity
\be
\cs\left( \S^{(i,ii)} \right) =0\label{ST},
\end{equation}
where the Slavnov-Taylor operator is given, for any functional $\cf$, by
\be
\cs\left( \cf \right) = \int d^{4}x \left( \pa_{\m}c \frac{\d\cf}{\d A_{\m}}  +    B \frac{\d\cf}{\d \bar{c}}      \right).
\end{equation}

\section{Photon Self-Energy}
In order to check the one-loop UV and IR behaviour of the actions (\ref{ACTION-gf}) and (\ref{ACTION-gfII}), one needs the corresponding Feynman rules. For the various propagators of the models only the bilinear part of the full actions is relevant. However, this is independent of $\q$ and thus the propagators are identical in both cases
\unitlength 1mm
\ba
\begin{picture}(30,8)
\put(-19,-7){
\parbox{4.2cm}{
\epsfig{figure=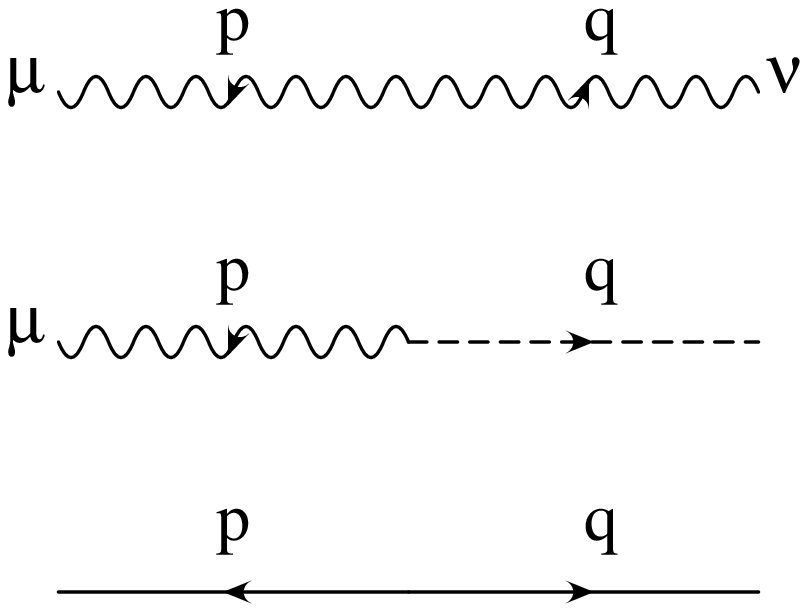,width=4.2cm}}
}
\end{picture}
\ti{G}^{AA}_{\m\n}(p)&=&~\frac{1}{(p^{2}+{\rm{i}}\e)}\left(g_{\m\n}-(1-\a)\frac{p_{\m}p_{\n}}{(p^{2}+{\rm{i}}\e)}\right),\\
\ti{G}^{AB}_{\m}(p)&=&~\frac{-{\rm{i}}p_{\m}}{(p^{2}+{\rm{i}}\e)},\\
\ti{G}^{\bar{c}c}(p) &=& \frac{-1}{(p^{2}+{\rm{i}}\e)},
\ea
with $p+q=0$. The action (\ref{actionII}) represents free Maxwell theory in the limit $\q\rightarrow 0$ . To first order in $\q$ the photon vertex reads:\\

\ba
\begin{picture}(30,10)
\put(-18,5){
\parbox{4.2cm}{
\epsfig{figure=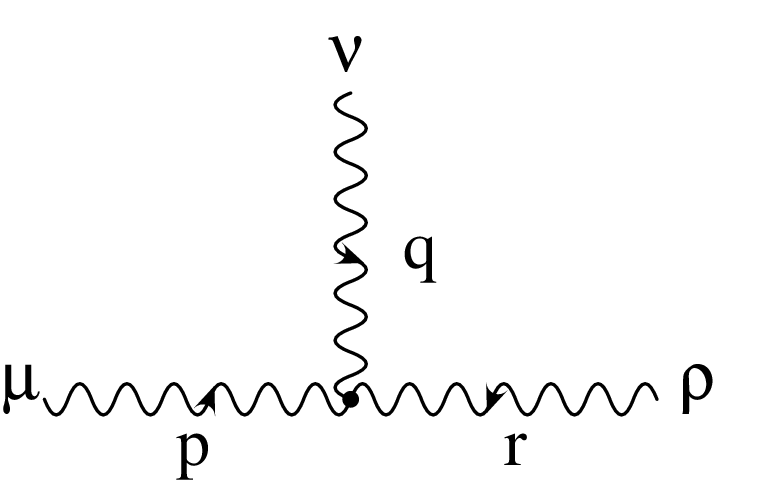,width=4.5cm}}}
\end{picture}
\ti{V}_{AAA}^{\m\n\r}(p,q,r)=-{\rm{i}}\theta_{\a\b}\OO^{\a\b\m\n\r}(p,q,r), \label{VER}
\ea
$ $\\
with
\begin{align}
\OO^{\a\b\m\n\r}(p,q,r)&=
 g^{\a\m}g^{\b\n}\big( (pr)q^{\r} -(qr)p^{\r} \big)
+g^{\a\n}g^{\b\r}\big( (qp)r^{\m} -(rp)q^{\m} \big)
+g^{\a\r}g^{\b\m}\big( (rq)r^{\n} -(pq)r^{\n} \big) \nonumber
\\
&+ g^{\a\m} \big( 
 ( g^{\n\r} (rq) - r^\n q^\r ) p^{\b}
-( g^{\n\r} (pq) - p^\n q^\r ) r^{\b}
-( g^{\n\r} (rp) - r^\n p^\r ) q^{\b}\big)  \nonumber
\\
&+ g^{\a\n} \big( 
 ( g^{\r\m} (pr) - p^\n r^\m ) q^{\b}
-( g^{\r\m} (qr) - q^\n r^\m ) p^{\b}
-( g^{\r\m} (pq) - p^\n q^\m ) r^{\b}\big)  \nonumber
\\
&+ g^{\a\r} \big( 
 ( g^{\m\n} (qp) - q^\m p^\n ) r^{\b}
-( g^{\m\n} (rp) - r^\m p^\n ) q^{\b}
-( g^{\m\n} (qr) - q^\m r^\n ) p^{\b}\big)  \nonumber
\\
&- g^{\m\n}\big(p^{\r} q^{\a} r^{\b} + q^{\r} p^{\a} r^{\b} \big)
 - g^{\n\r}\big(q^{\m} r^{\a} p^{\b} + r^{\m} q^{\a} p^{\b} \big)
 - g^{\r\m}\big(r^{\n} p^{\a} q^{\b} + p^{\n} r^{\a} q^{\b} \big)\;,
\end{align}
and $p+q+r=0$. In the linear gauge the ghost is Abelian and does not couple to the gauge-field. In the nonlinear gauge the action (\ref{ACTION-gfII}) leads to the following interactions:
\ba
\begin{picture}(20,8)
\put(-19,-17){
\parbox{4.2cm}{
\epsfig{figure=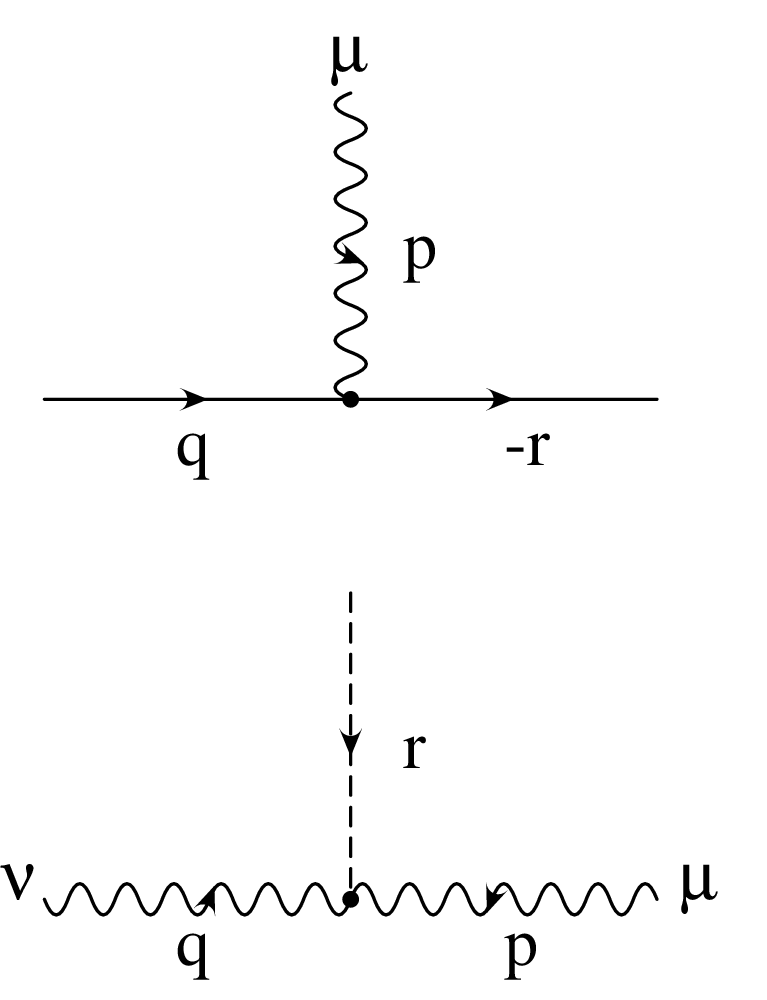,width=4.2cm}}
}
\end{picture}
&&\nonumber\\
&&\ti{V}_{A\bar{c}c}^{\m}(p,q,r)=-{\rm{i}}\theta_{\a\b}\left(\frac{1}{2}q^{2}r^{\b}g^{\m\a}+p^{\a}r^{\b}q^{\m}  \right),  \nonumber\\
&&\nonumber\\
&&\nonumber\\
&&\nonumber\\
&& \ti{V}_{AAB}^{\m\n}(p,q,r)=\nonumber\\ && \;\;\;\;\;\;\;\;\theta_{\a\b} \left(-\frac{1}{2}g^{\a\m}g^{\b\n}(pr)                 
+\frac{1}{2}g^{\a\m}g^{\b\n}(qr) 
-g^{\a\m}q^{\b}r^{\n}  
-g^{\a\n}p^{\b}r^{\m}       \right),
\ea
with $p+q+r=0$.

 As usual, for each independent loop momentum $k_{i}$ we have the integration operator $\frac{\hbar}{{\rm{i}}}\int \frac{d^{4}k_{i}}{(2\p)^{4}}$ and momentum conservation for the external momenta $p_{i}$ leading to a factor $(2\p)^{4}\d(\S p_{i})$. Each closed ghost line contributes a factor $-1$.\\

 Before doing explicit one loop analysis we want to stress that the Ward identity (\ref{ST}) implies that the radiative corrections to the photon propagator must be transversal
\be
p_{\m}\P^{\m\n}(p)=0.\label{trans}
\end{equation}
Furthermore, (\ref{trans}) implies that the radiative corrections up to first order in $\q$ must vanish (there are no $\q$-independent interactions)
\be
\P^{\m\n}(p)=0\label{Null},\;\;\;\;\;\;\;\;\mbox{(order $\q$).}
\end{equation}
The radiative corrections up to second order in $\q$ are restricted in form by
\ba
\P^{\m\n}(p)&=&\left(g^{\m\n}p^{2}-p^{\m}p^{\n}\right)\P^{(i)}(p)
+\ti{p}^{\m}\ti{p}^{\n}\P^{(ii)}(p)\nonumber\\&&
+\left(\ptt^{\m}p^{\n}+\ptt^{\n}p^{\m}+g^{\m\n}\ti{p}^{2}+p^{2}\q^{\m}_{\,\,\,\s}\q^{\n\s}   \right) \P^{(iii)}(p)\;\;\;\;\;  (\mbox{order $\q$}^{2}).\label{form1}
\ea
where $\ti{p}_{\m}=\q_{\m\n}p^{\n}$ and $\ptt_{\m}=\q_{\m\n}\q^{\n\r}p_{\r}$. In (\ref{form1}) we used that $\ti{p}$ is orthogonal to $p$ and $\ptt$ and that $p$ and $\ptt$ are independent.
Notice that due to the negative dimension of $\q$, (\ref{form1}) indicates the presence of (divergent) Feynman graphs with 6 powers of $p$ in $\P^{\m\n}(p)$. Since the bilinear part of the action (\ref{action}) is the ordinary one of Maxwell theory, such a term will be non-renormalizable.  
\begin{figure}
\begin{center}
\epsfig{figure=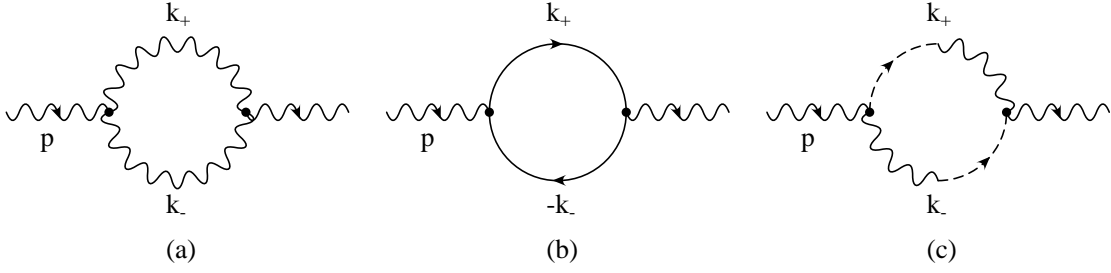,width=15cm}
\caption{Self-energy graphs}
\end{center}
\end{figure}
 
 In the following we will explicitly perform the one loop analysis of the photon self-energy. Since all vertices are linear in $\q$ the first contribution is proportional to $~\q^{2}$. In the linear gauge the only contributing graph is shown in fig.$\,$1.a. In the nonlinear gauge we have interacting ghost and multiplier fields and thus find contributions from all three graphs shown in fig.$\,$1. In fact we should also consider the tadpole graph emerging from the Seiberg-Witten map to second order in $\q$ via a 4-legged photon interaction. However, the tadpole graph is identically zero, because there is no mass in the theory. Using the above Feynman rules one calculates the following expression for the photon self-energy with an internal photon line
\be
\P^{(a),\m\n}(p)=\frac{\hbar}{2{\rm{i}}} \int \frac{d^{4}k}{(2\p)^{4}}\ti{V}_{AAA}^{\m\r\s}(p,-k_{+},-k_{-})\ti{V}^{\n\k\l}_{AAA}(-p,k_{-},k_{+})\ti{G}^{AA}_{\r\l}(-k_{+})\ti{G}^{AA}_{\k\s}(k_{-}),
\end{equation}
where $k_{+}=\frac{p}{2}+k$, $k_{-}=\frac{p}{2}-k$. The relevant integrals are evaluated in the Appendix. We find
\ba
\P^{(a),\m\n}(p)&=&\frac{\hbar}{(4\p)^{2}\ve}\Big(
-\frac{1}{8}(p^{2})^2\q^2    \left(g^{\m\n}p^{2}-p^{\m}p^{\n}\right)+
\frac{1}{10}\ti{p}^{2}p^{2} \left(g^{\m\n}p^{2}-p^{\m}p^{\n}\right)+
\frac{1}{30}(p^{2})^2\ti{p}^{\m}\ti{p}^{\n}      \nonumber\\&&
+\frac{1}{4} (p^{2})^2\left( \ti{\ti{p}}^{\m}p^{\n}+\ti{\ti{p}}^{\n}p^{\m}+g^{\m\n}\ti{p}^{2}+p^{2}\q^{\m}_{\,\,\,\s}\q^{\n\s}  \right)  \Big)+O(1).\label{loop1}
\ea
Notice that (\ref{loop1}) satisfies the transversality condition (\ref{form1}). For the graph (b) the integral reads
\be
\P^{(b),\m\n}(p)=-\frac{\hbar}{{\rm{i}}} \int \frac{d^{4}k}{(2\p)^{4}}\ti{V}_{A\bar{c}c}^{\m}(p,-k_{+},-k_{-})\ti{V}^{\n}_{A\bar{c}c}(-p,k_{-},k_{+})\ti{G}^{\bar{c}c}(-k_{+})\ti{G}^{\bar{c}c}(k_{-}).
\end{equation}
We find
\be
\P^{(b),\m\n}(p)=-\frac{\hbar}{60(4\p)^{2}\ve}\Big(
\frac{1}{4}(p^{2})^2\ti{p}^{2}g^{\m\n}+
p^{2}\ti{p}^{2}p^{\m}p^{\n}+
\frac{1}{2}(p^{2})^2\ti{p}^{\m}\ti{p}^{\n}\Big)+O(1).\label{loop2}
\end{equation}
For the graph (c) we write
\be
\P^{(c),\m\n}(p)=\frac{\hbar}{{\rm{i}}} \int \frac{d^{4}k}{(2\p)^{4}}\ti{V}_{AAB}^{\m\r}(p,-k_{+},-k_{-})\ti{V}^{\n\s}_{AAB}(-p,k_{-},k_{+})\ti{G}^{AB}_{\r}(-k_{+})\ti{G}^{AB}_{\s}(k_{-}),
\end{equation}
and find
\be
\P^{(c),\m\n}(p)=\frac{\hbar}{60(4\p)^{2}\ve}\Big(
\frac{1}{4}(p^{2})^2\ti{p}^{2}g^{\m\n}+
p^{2}\ti{p}^{2}p^{\m}p^{\n}+
\frac{1}{2}(p^{2})^2\ti{p}^{\m}\ti{p}^{\n}\Big)+O(1).\label{loop3}
\end{equation}
One sees that the above divergent contributions from the ghost graph (b) and the multiplier-photon graph (c) cancel identically. This means that the choice of linear or non-linear gauge leaves the renormalization invariant. Furthermore, we would like to stress that the radiative correction (\ref{loop1}) is independent of $\a$, which shows that our result is gauge-independent. The reason for this is that the vertex (\ref{VER}) is transversal, $p_{\m}\ti{V}_{AAA}^{\m\n\r}(p,q,r)=0$.

\section{Higher Derivative Action}

In the previous section we have shown that the radiative corrections
to the photon self-energy produce divergent terms involving two orders of $\q$
and six orders of $p$. These terms cannot be absorbed into counterterms to
the initial action (\ref{actionII}), which thus is perturbatively 
non-renormalizable. We interpret this problem as a hint to extend the
classical action. 

The extension to (\ref{actionII}) must be invariant under Lorentz
transformations and the Abelian gauge transformations (\ref{gaugeII}).
There are many possibilities to write down the same terms. A
generalization to non-Abelian models suggests however to use the field
strengths $F_{\mu\nu}$ and $\tilde{F}_{\mu\nu} :=
\theta_\mu^{~\;\alpha} F_{\alpha\nu}$ as well as their derivatives using
the operators $\partial_\mu$ and $\tilde{\partial}_\mu :=
\theta_\mu^{~\;\alpha}\partial_\alpha$ as building blocks. Thus we
have the following tensors of dimension 2 at disposal:
\begin{align}
F_{\mu\nu}&~,  
&
\tilde{F}'_{\mu\nu} &:= \partial_\mu \tilde{\partial}^\alpha F_{\alpha\nu}~, 
&
\tilde{F}''_{\mu\nu} &:= \tilde{\partial}^\alpha \partial_\mu F_{\alpha\nu}~, 
\nonumber\\
\tilde{F}_{\mu\nu\rho\sigma} &:= \partial_\mu
\partial_\nu \tilde{F}_{\rho\sigma}~, 
&
\tilde{F}'_{\mu\nu\rho\sigma} &:= \partial_\mu
\tilde{\partial}_\nu F_{\rho\sigma}~, 
&
\tilde{F}''_{\mu\nu\rho\sigma} &:= 
\tilde{\partial}_\mu \partial_\nu F_{\rho\sigma}~, 
\nonumber\\
\tilde{F}_{\kappa\lambda\mu\nu\rho\sigma} &:= 
\theta_{\kappa\lambda} \partial_\mu \partial_\nu F_{\rho\sigma}~.
\end{align}
The abelian case is degenerate; we have
$\tilde{F}'_{\mu\nu}=\tilde{F}''_{\mu\nu}$
and
$\tilde{F}'_{\mu\nu\rho\sigma}=\tilde{F}''_{\nu\mu\rho\sigma}$.

The most general gauge and Lorentz invariant extension to (\ref{action})
of dimension 4 with two $\theta$'s is\footnote{Observe that
$\int d^4x\;\tilde{F}_{\mu\nu\rho\sigma}\tilde{F}'{}^{\mu\nu\rho\sigma}= 
\int d^4x\;\tilde{F}_{\mu\nu\rho\sigma}\tilde{F}''{}^{\mu\nu\rho\sigma}= 
\int d^4x\;\tilde{F}'_{\mu\nu\rho\sigma}\tilde{F}''{}^{\mu\nu\rho\sigma}=
0$.}$^,$\footnote{We may add that all terms involving tensorial combinations linear in $\q$ are either identically zero or zero after integration (topological terms).}
\begin{align}
\Sigma_{ext} = \int \!\! d^4x \Big( 
\frac{1}{4 g_1^2} \tilde{F}'_{\mu\nu}\tilde{F}'{}^{\mu\nu} +
\frac{1}{4g_2^2} 
\tilde{F}_{\mu\nu\rho\sigma}\tilde{F}^{\mu\nu\rho\sigma} 
+ \frac{1}{4 g_3^2}
\tilde{F}'_{\mu\nu\rho\sigma}\tilde{F}'{}^{\mu\nu\rho\sigma} - 
\frac{\mathrm{sign}(\theta_{\alpha\beta}\theta^{\alpha\beta})}{4 g_4^2}
\tilde{F}_{\kappa\lambda\mu\nu\rho\sigma}
\tilde{F}^{\kappa\lambda\mu\nu\rho\sigma} \Big)\,.\label{ext0}
\end{align}
The signs are chosen such that the highest time derivatives are
positive, \emph{i.e.} that the action is bounded from below. This requires for the second term 
\[
H_2^{ij} = \theta^{i0}\theta^{j0} + \sum_{k\neq 0} ( \theta^{k0}\theta^{k0}
\delta^{ij} - \theta^{ki}\theta^{kj}) \geq 0~,\qquad i,j\neq 0~.
\]
For example, the case where the only non-vanishing commutators are 
$[x^0,x^3]=\mathrm{i} \Theta_1$ and $[x^1,x^2]=\mathrm{i} \Theta_2$, requires 
$|\Theta_1| \geq |\Theta_2|$.

We remark that the action (\ref{ext0}) is bilinear in the gauge field. Therefore the photon propagator is changed, thus changing the whole scheme of quantization. The treatment of higher derivative actions have been investigated in the literature, see \emph{e.g}\ \cite{Buchbinder:1992rb} and references therein. 

Here we choose to view $\q$ as a {\it constant external field}, thus consider the photon propagator as unchanged and the action (\ref{ext0}) as new vertices of type $AA\q\q$. In this sense (\ref{loop1}) represents the proper one-loop radiative correction to the coupling constants in (\ref{ext0}). 

The result of our one-loop calculation was the independence from the
gauge parameter. This implies that we can have the special solution of a
single coupling constant. From (\ref{loop1}) we conclude the reduction
to the following extended action:
\begin{align}
\Sigma_{ext}^{red} = \frac{1}{4 g^2(\ve)} \int \!\! 
d^4x \Big( \frac{2}{15} \tilde{F}'_{\mu\nu}\tilde{F}'{}^{\mu\nu} 
+ 
\tilde{F}_{\mu\nu\rho\sigma}\tilde{F}^{\mu\nu\rho\sigma} 
+ \frac{1}{5}
\tilde{F}'_{\mu\nu\rho\sigma}\tilde{F}'{}^{\mu\nu\rho\sigma} - 
\frac{1}{4}\tilde{F}_{\kappa\lambda\mu\nu\rho\sigma}
\tilde{F}^{\kappa\lambda\mu\nu\rho\sigma} \Big)\,.
\label{ext}
\end{align}
with
\begin{equation}
g^2(\varepsilon) = g_0^2 (1 + \frac{g_0^2\hbar}{4 (4\pi)^2 \varepsilon} 
+ O(g_0^4\hbar^2))~. \label{cc}
\end{equation}
The highest time derivatives in (\ref{ext}) are $H^{ij} 
(\partial_0^3 A_i) (\partial_0^3 A_j)$ with
\begin{equation}
H^{ij} = \frac{17}{60} \theta^{i0}\theta^{j0} + \sum_{k\neq 0} 
\frac{1}{10} \theta^{k0}\theta^{k0} \delta^{ij} 
+ \frac{1}{4} \Big(\sum_{l>k \neq 0} \theta^{kl}\theta^{kl} \delta^{ij}
-\sum_{k\neq 0} \theta^{ki}\theta^{kj}\Big) > 0~,
\end{equation}
i.e.\ for any $\theta$ the reduced extended action is bounded from below. The result
(\ref{cc}) tells us that the extended action is not asymptotically free.

Applying the Seiberg-Witten map in the opposite sense, the action
(\ref{ext}) should arise from some noncommutative action 
$\hat{\Sigma}_{ext}$. Gauge invariance leads immediately to the solution 
\begin{align}
\hat{\Sigma}_{ext} = \frac{1}{4 g^2} \int &
\Big( 
\frac{2}{15} \Big(\beta_1 
\hat{\tilde{F}}{}'_{\mu\nu}\hat{\tilde{F}}{}'{}^{\mu\nu} 
+(1{-}\beta_1)\beta_2 
\hat{\tilde{F}}{}''_{\mu\nu}\hat{\tilde{F}}{}''{}^{\mu\nu} 
+(1{-}\beta_1)(1{-}\beta_2) 
\hat{\tilde{F}}{}'_{\mu\nu}\hat{\tilde{F}}{}''{}^{\mu\nu} 
\Big) \nonumber
\\
&+  \frac{1}{5}\Big( \beta_3
\hat{\tilde{F}}{}'_{\mu\nu\rho\sigma}
\hat{\tilde{F}}{}'{}^{\mu\nu\rho\sigma} 
+(1{-}\beta_3) \hat{\tilde{F}}{}''_{\mu\nu\rho\sigma}
\hat{\tilde{F}}{}''{}^{\mu\nu\rho\sigma}  
\Big) + 
\hat{\tilde{F}}_{\mu\nu\rho\sigma}\hat{\tilde{F}}{}^{\mu\nu\rho\sigma} 
- \frac{1}{4}\hat{\tilde{F}}_{\kappa\lambda\mu\nu\rho\sigma}
\hat{\tilde{F}}{}^{\kappa\lambda\mu\nu\rho\sigma} \nonumber\\&
+  \gamma_1 \hat{\tilde{F}}_{\mu\nu\rho\sigma}
\hat{\tilde{F}}{}'{}^{\mu\nu\rho\sigma} 
+\gamma_2 \hat{\tilde{F}}_{\mu\nu\rho\sigma}
\hat{\tilde{F}}{}''{}^{\mu\nu\rho\sigma}
+\gamma_3 
\hat{\tilde{F}}{}'_{\mu\nu\rho\sigma}
\hat{\tilde{F}}{}''{}^{\mu\nu\rho\sigma}\Big)\,,
\label{extI}
\end{align}
for $0\leq \beta_i\leq 1$, where
\begin{align*}
\hat{\tilde{F}}_{\mu\nu} &:= \theta_\mu^{~\;\alpha} \hat{F}_{\alpha\nu}~,
& 
\hat{\tilde{D}}_\mu &:= \theta_\mu^{~\;\alpha} \hat{D}_\alpha~,
\\
\hat{\tilde{F}}{}'_{\mu\nu} &:= \hat{D}_\mu \hat{\tilde{D}}{}^\alpha 
\hat{F}_{\alpha\nu}~, 
&
\hat{\tilde{F}}{}''_{\mu\nu} &:= \hat{\tilde{D}}{}^\alpha \hat{D}_\mu 
\hat{F}_{\alpha\nu}~, 
\\
\hat{\tilde{F}}_{\mu\nu\rho\sigma} &:= \hat{D}_\mu
\hat{D}_\nu \hat{\tilde{F}}_{\rho\sigma}~, 
&
\hat{\tilde{F}}{}'_{\mu\nu\rho\sigma} &:= \hat{D}_\mu
\hat{\tilde{D}}_\nu \hat{F}_{\rho\sigma}~, 
&
\hat{\tilde{F}}{}''_{\mu\nu\rho\sigma} &:= 
\hat{\tilde{D}}_\mu \hat{D}_\nu \hat{F}_{\rho\sigma}~, 
\\
\hat{\tilde{F}}_{\kappa\lambda\mu\nu\rho\sigma} &:= 
\theta_{\kappa\lambda} \hat{D}_\mu \hat{D}_\nu \hat{F}_{\rho\sigma}~.
\end{align*}
Note that the action (\ref{extI}) leads, after applying the
Seiberg-Witten map, to an action containing infinitely many additional terms with finitely many free coefficients. The fact that the renormalization of the self-energy radiative correction puts restrictions on the relative weights of possible counterterms for the Green's function with three external legs provides us with a strong test of the model. We will address this question in a
forthcoming paper \cite{NEXT}.

\section{Conclusion}

We have analyzed the $\theta$-expanded noncommutative U(1) Yang-Mills
theory as a \emph{perturbative quantum field theory}. As expected from
the power-counting behaviour the Yang-Mills action $\int
\hat{F}_{\mu\nu} \hat{F}^{\mu\nu}$ is not renormalizable in this
setting. We singled out the \emph{unique} extended action for which
the one-loop photon propagator is renormalizable.

Lorentz and gauge invariance allow for four different extension terms
with arbitrary coefficients (coupling constants). Our one-loop
calculations reduce this freedom to a single coupling constant, due to
two not anticipated facts: the independence from the gauge parameter
and from linear versus non-linear gauge. 

We are thus led to ask whether there is a meaning in the relative
weights of the extension terms. We recall in this respect the
remarkable agreement of all three relative signs, which ensures that
the action is bounded from below also for large momenta $|p_0| \gg
|\theta|^{-\frac{1}{2}}$. It would be interesting to investigate whether
$\theta$-expanded noncommutative QED leads to the same weights.

It is obvious that the extension we derived is only valid to lowest
order in $\theta$. The new vertices lead to non-renormalizable
divergences which give rise to more and more 
extension terms. Hence the action makes sense only as the lowest-order
parts of an effective theory. 

There are two ways a factor $\theta$ can arise in the
$\theta$-expansion of the noncommutative Yang-Mills action: in the
form $\theta p A$ via the Seiberg-Witten map and in the form $\theta
p^2$ via the deformation product and possibly higher-order
Seiberg-Witten terms. This leads to a field strengh of structure
\be 
\sum_{\sigma,\delta} \Big( x_{\sigma\delta} (pA)(\theta p A)^\sigma 
(\theta p^2)^\delta + y_{\sigma\delta}
(\theta p^2) A^2 (\theta p A)^\sigma (\theta p^2)^\delta\Big),
\end{equation}
 with the
very important restriction $x_{\delta 0}=0$ for all $\delta$. A Feynman
graph with $E$ external $A$-lines and $L$ loops has then the structure
$p^E \theta^{E-2} (\theta p^2)^{2L+\Delta}$, where $\Delta$ is the
total number of deformations $\delta$ in the vertices of the graph. It
follows that, in principle, divergences in coefficients to
factors $\theta p^2$ from integrated higher loop graphs can be
absorbed by terms with a higher $\Delta$ in the tree action.

But this mechanism does not work for $E=2$ and $L=0$; in this case the
tree action has $\Delta\equiv 0$. In other words, there is no chance
that the photon propagator corrections are renormalizable. This is why
we are forced to add to the tree action something with $\Delta=2$ in
order to compensate the $L=1$ divergences. It is also clear that for
compensating higher and higher loop graphs we need additional terms
with arbitrarily large $\Delta$ in the tree action. In some sense this
makes the tree action more symmetric with respect to the power of
$\theta p^2$.

We would like to suggest the following interpretation of the extra
terms to the Yang-Mills tree action. There is a remarkable structural
asymmetry  between the product of fields in NCYM (which contains
arbitrarily many factors $\theta p^2$ in the $\star$-product) and the
trace where the $\star$-product is reduced to the ordinary
product. The extra terms we found restore the symmetry in \emph{deforming
the trace as well}. Differentiations in the scalar product are not
unfamiliar, for instance, the Sobolev norm of $f\in H_s$ is given by 
\be
\|f\|^2_{H_s} \equiv \langle f,f\rangle_{H_s} = 
\int \! dx\; \Big(|f(x)|^2 + \sum_{\alpha, ~1\leq |\alpha| \leq s} 
a_\alpha |\partial^\alpha_x  f(x)|^2\Big)~,\label{label}
\end{equation}
where $\alpha$ is a multi-index. 

In this context, we have derived in this paper the necessity to
replace the $L^2$ scalar product $\langle
\hat{F},\hat{F}\rangle_{L^2}$ for the field strength by the $H_\infty$
scalar product $\langle \hat{F},\hat{F}\rangle_{H_\infty}$. Since the
coordinate $x$ has a dimension in physics, the derivatives must be
accompanied by a dimensionful parameter $\theta$. Of
course, this scalar product must be gauge invariant, therefore we must
take covariant derivatives in the Sobolev norm instead of partial
derivatives. The dependence of the scalar product on the gauge field
is very natural in the framework of noncommuatative geometry, where
actions are built out of the covariant Dirac operator
\cite{Connes:1996gi}. Moreover, the boundedness of the action from below
gives certain restrictions on the pre-factors $a_\alpha$ of the different
combinations of $\theta^{\rho\sigma}$ and $\hat{D}_\mu$. 
We would like to stress that in the commutative limit $\theta\to 0$
the $H_\infty$ scalar product reduces to the standard $L^2$ scalar product.

Hence, the big quest is to find the true $H_\infty$ scalar product (the prefactors $a_{\a}$ in (\ref{label}))
which makes the $\theta$-expanded Yang-Mills action
renormalizable. In this paper we have succeeded to derive the first
correction to the $L^2$ scalar product -- our result (\ref{extI}).  
We may speculate whether the relative weights we computed can serve as a
hint in which direction to search for a closed form of the
renormalizable $H_\infty$ scalar product. 

We may also speculate whether this renormalizable $H_\infty$ scalar
product also solves the UV/IR-mixing problem of the
$\theta$-unexpanded Yang-Mills action on noncommutative
$\mathbb{R}^4$. We recall that the $\theta$-expansion is free of
infrared divergences but UV non-renormalizable whereas the unexpanded
version is IR non-renormalizable \cite{Matusis:2000jf}\footnote{We
  refer to \cite{Chepelev:2000hm} for the power-counting behaviour of
  field theories on noncommuative $\mathbb{R}^D$.}. This can be
interpreted as a hint to extend the Yang-Mills action also in the
$\theta$-unexpanded setting, and one could speculate if the solution is to substitute
the ordinary scalar product with the $H_\infty$
scalar product which is renormalizable via $\theta$-expansion. Thus
our result could be valuable also for the $\theta$-undeformed
framework. We would like to remark that the $H_\infty$ scalar product
leads to a $\theta$-dependend photon propagator and could make contact
with a different approach \cite{Cho:2000sg} to the noncommutative
$\mathbb{R}^4$.

Finally let us mention that the extended action leads to a modified
wave equation for the photon already on tree-level. Since the
modification is of the order $|\theta|^2 |p|^4$, and if we assume
$|\theta|^{1/2}$ to be of the order of the Planck length, there can 
be observable consequences only for extremely high-energetic (cosmological)
phenomena.

\section{Acknowledgement}

The authors would first of all like to thank Julius Wess for giving us the initial idea as well as for enlightening discussions. Also we thank Martin Ertl for his help. The very involved calculations found in this paper were performed using his fantastic \emph{Mathematica}$^{TM}$ package ``Index''. Furthermore, we would like to thank Harald Grosse, Karl Landsteiner, Stefan Schraml and Raymond Stora for fruitful discussions.

\allowdisplaybreaks[2]
\begin{appendix}
\section{Integrals}

We use Zimmermann's $\epsilon$-trick \cite{Zimmermann:1969jj} and replace $\frac{1}{k^2 
+\mathrm{i}\epsilon}=\frac{1}{k_0^2-\vec{k}^2 +\mathrm{i}\epsilon}$ by 
$\frac{1}{k_0^2-\vec{k}^2 +\mathrm{i}\epsilon \vec{k}^2}$. Then, 
\begin{align}
P(k,p) &= \lim_{\epsilon \to 0} 
\frac{1}{
((\frac{p_0}{2}{-}k_0)^2 - (\frac{\vec{p}}{2}{-}\vec{k})^2
  +\mathrm{i} \epsilon (\frac{\vec{p}}{2}{-}\vec{k})^2)
((\frac{p_0}{2}{+}k_0)^2 - (\frac{\vec{p}}{2}{+}\vec{k})^2
  +\mathrm{i} \epsilon (\frac{\vec{p}}{2}{+}\vec{k})^2)}
\nonumber 
\\
& = \lim_{\epsilon \to 0} \int_0^1 \!\! dx\;
\frac{(\epsilon'{-}\mathrm{i})^2}{\{
(\epsilon'{-}\mathrm{i})(k_0^2 + (1{-}2x) k_0 p_0 + \frac{1}{4} p_0^2)
- (\epsilon'{-}\mathrm{i})(1{-}\mathrm{i}\epsilon)(
\vec{k}^2 + (1{-}2x)\vec{k}\vec{p}+ \frac{1}{4}\vec{p}\,^2) \}^2}~.
\label{denom}
\end{align}
For $\epsilon' < \epsilon$ we have $\mathrm{Re}(\{\dots\})> 0$ in the
denominator of (\ref{denom}). We use analytic regularization \cite{Speer:1975gj} to write
$\frac{(\epsilon'-\mathrm{i})^2}{\{\dots\}^2} \to 
\frac{\mu^{2\varepsilon}(\epsilon'-\mathrm{i})^{2+\varepsilon}}{
\{\dots\}^{2+\varepsilon}}$ and rewrite $P(k,p)$ in terms of the
Schwinger parameter $\alpha$:
\begin{align}
P(k,p) &\to 
\lim_{\epsilon \to 0,~\epsilon'<\epsilon}
\frac{\mu^{2\varepsilon}(\epsilon'{-}\mathrm{i})^{2+\varepsilon}}{
\Gamma(2+\varepsilon)}
\int_0^1 dx \int_0^\infty d\alpha\;
\alpha^{1+\varepsilon} \nonumber
\\
&\hspace*{9em} \times
e^{-(\epsilon'{-}\mathrm{i})\alpha 
(k_0^2 + k_0 q_0 - \vec{k}\vec{q} + \frac{1}{4}p_0^2) 
- (\epsilon-\epsilon'+\mathrm{i}
+\mathrm{i}\epsilon\epsilon')\alpha (\vec{k}^2
+ \frac{1}{4}\vec{p}\,^2)}
\Big|_{{q_0=(1{-}2x)p_0 \atop 
\vec{q}=(1{-}2x)(1-\mathrm{i}\epsilon)\vec{p}}}~.
\end{align}
Factors $k_\mu$ in the numerator can now be obtained by
differentiation with respect to $q$. For $\varepsilon > 0$ the
various integrations can be performed and yield
\begin{align}
\label{int0}
\lim_{\epsilon \to 0} &
\int \frac{d^4k}{(2\pi)^4}\;
\frac{1}{((\frac{p}{2}-k)^2+\mathrm{i}\epsilon)
((\frac{p}{2}+k)^2+\mathrm{i}\epsilon)}
\\*
\nonumber
&=  \frac{\mathrm{i}}{ (4\pi)^2} \Big(\frac{1}{\varepsilon} + 
\ln\Big(\frac{\mu^2}{p^2}\Big)\Big) + \frac{\mathrm{i}}{(4\pi)^2} 
\Big({-}1 + (\gamma {+} \ln 4 {+} \psi(\tfrac{3}{2}))\Big)  + O(\varepsilon)~,
\\[1ex]
 \label{int1}
\lim_{\epsilon\to 0}& \int \frac{d^4k}{(2\pi)^4}\;\frac{k_\mu  k_\nu}{
((\frac{p}{2}-k)^2+\mathrm{i}\epsilon)
((\frac{p}{2}+k)^2+\mathrm{i}\epsilon)}
\\*
& = \frac{\mathrm{i}}{12 (4\pi)^2}\Big(\frac{1}{\varepsilon} + 
\ln\Big(\frac{\mu^2}{p^2}\Big)\Big)
( p_\mu p_\nu - g_{\mu\nu} p^2)  \nonumber
\\*
& + \frac{\mathrm{i}}{(4\pi)^2} \Big(g_{\mu\nu} p^2
(\tfrac{1}{12}-\tfrac{1}{12}(\gamma{+}\ln4{+}\psi(\tfrac{5}{2}))) 
+ p_\mu p_\nu (\tfrac{23}{36}
-\tfrac{1}{4}(\gamma{+}\ln4{+}\psi(\tfrac{3}{2}))) 
\Big) +O(\varepsilon) ~, \nonumber
\\[1ex]
\lim_{\epsilon\to 0} & 
\int  \frac{d^4k}{(2\pi)^4}\;\frac{k_\kappa k_\lambda k_\mu  k_\nu}{
((\frac{p}{2}-k)^2+\mathrm{i}\epsilon)
((\frac{p}{2}+k)^2+\mathrm{i}\epsilon)} 
\\*
&= \frac{\mathrm{i}}{240(4\pi)^2}
\Big(\frac{1}{\varepsilon}+\ln\Big(\frac{\mu^2}{p^2}\Big)\Big)
\Big( (p^2)^2 T^0_{\kappa\lambda\mu\nu}
-p^2 T^2_{\kappa\lambda\mu\nu} + 3 T^4_{\kappa\lambda\mu\nu} \Big)
\nonumber
\\*
& + \frac{\mathrm{i}}{(4\pi)^2} \Big(
(p^2)^2 T^0_{\kappa\lambda\mu\nu}(p)
(-\tfrac{1}{240}+\tfrac{1}{240}(\gamma{+}\ln 4{+}\psi(\tfrac{7}{2})))
+ p^2 T^2_{\kappa\lambda\mu\nu}(p)
(-\tfrac{77}{1200}+\tfrac{1}{48}(\gamma{+}\ln 4{+}\psi(\tfrac{5}{2})))
\nonumber
\\*
& \qquad\quad +T^4_{\kappa\lambda\mu\nu}(p)
(\tfrac{481}{1200}-\tfrac{3}{16}(\gamma{+}\ln 4{+}\psi(\tfrac{3}{2})))
\Big) + O(\varepsilon)~, \nonumber
\\[1ex]
\lim_{\epsilon\to 0} & 
\int \frac{d^4k}{(2\pi)^4}\;\frac{k_\kappa k_\lambda k_\mu  k_\nu
k_\rho  k_\sigma}{
((\frac{p}{2}-k)^2+\mathrm{i}\epsilon)
((\frac{p}{2}+k)^2+\mathrm{i}\epsilon)} \label{int3} 
\\*
=& \frac{\mathrm{i}}{6720 (4\pi)^2}\Big(\frac{1}{\varepsilon}
{+}\ln\Big(\frac{\mu^2}{p^2}\Big)\Big) \Big({-} (p^2)^3
\,T_{\kappa\lambda\mu\nu\rho\sigma}^0(p)
+ (p^2)^2
\,T_{\kappa\lambda\mu\nu\rho\sigma}^2(p)  
\nonumber
\\*
&\qquad\quad 
- 3 p^2 \,T_{\kappa\lambda\mu\nu\rho\sigma}^4(p)
+ 15 T_{\kappa\lambda\mu\nu\rho\sigma}^6(p) \Big) \nonumber
\\*
& ~~ + \frac{\mathrm{i}}{(4\pi)^2} \Big( (p^2)^3
\,T_{\kappa\lambda\mu\nu\rho\sigma}^0(p) ( \tfrac{1}{6720}-
\tfrac{1}{6720}(\gamma{+}\ln 4{+}\psi(\tfrac{9}{2})))  
\nonumber
\\*
&~~
+ (p^2)^2\,T_{\kappa\lambda\mu\nu\rho\sigma}^2(p) ( \tfrac{2501}{705600}-
\tfrac{1}{960}(\gamma{+}\ln 4{+}\psi(\tfrac{7}{2}))) 
+ p^2\,T_{\kappa\lambda\mu\nu\rho\sigma}^4(p) 
( -\tfrac{3349}{78400}+
\tfrac{1}{64}(\gamma{+}\ln 4{+}\psi(\tfrac{7}{2})))  
\nonumber
\\*
&~~
+ T_{\kappa\lambda\mu\nu\rho\sigma}^6(p) ( \tfrac{7597}{47040}-
\tfrac{5}{64}(\gamma{+}\ln 4{+}\psi(\tfrac{7}{2})))\Big)
+ O(\varepsilon)~. \nonumber
\end{align}  
Here we have introduced the totally symmetric momentum tensors
\begin{align*}
T_{\kappa\lambda\mu\nu}^0(p) &:= \frac{1}{2!2!2!} \sum_{\pi \in
S(\kappa\lambda\mu\nu)} 
g_{\pi(\kappa)\pi(\lambda)}\,g_{\pi(\mu)\pi(\nu)}~,
\\*
T_{\kappa\lambda\mu\nu}^2(p) &:= \frac{1}{2!2!} \sum_{\pi \in
S(\kappa\lambda\mu\nu)} 
g_{\pi(\kappa)\pi(\lambda)}\,p_{\pi(\mu)} p_{\pi(\nu)}~,
\\*
T_{\kappa\lambda\mu\nu}^4(p) &:= p_\kappa p_\lambda p_\mu p_\nu ~,
\\
T_{\kappa\lambda\mu\nu\rho\sigma}^0(p) &:= 
\frac{1}{2!2!2!3!} \sum_{\pi \in
S(\kappa\lambda\mu\nu\rho\sigma)} 
g_{\pi(\kappa)\pi(\lambda)}\,g_{\pi(\mu)\pi(\nu)}\,
g_{\pi(\rho)\pi(\sigma)}~,
\\*
T_{\kappa\lambda\mu\nu\rho\sigma}^2(p) &:= 
\frac{1}{2!2!2!2!} \sum_{\pi \in
S(\kappa\lambda\mu\nu\rho\sigma)} 
g_{\pi(\kappa)\pi(\lambda)}\,g_{\pi(\mu)\pi(\nu)} \,p_{\pi(\rho)} 
p_{\pi(\sigma)}~,
\\*
T_{\kappa\lambda\mu\nu\rho\sigma}^4(p) &:= 
\frac{1}{2!4!} \sum_{\pi \in
S(\kappa\lambda\mu\nu\rho\sigma)} 
g_{\pi(\kappa)\pi(\lambda)} \,p_{\pi(\mu)} p_{\pi(\nu)} p_{\pi(\rho)} 
p_{\pi(\sigma)}~,
\\*
T_{\kappa\lambda\mu\nu\rho\sigma}^6(p) &:= 
p_\kappa p_\lambda p_\mu p_\nu p_\rho p_\sigma~,
\end{align*}
where $S(\mu_1\dots \mu_n)$ is the set of permutations of
the indices $\mu_1\dots \mu_n$. Let us finally mention that the
divergent parts of the above integrals (\ref{int1})-(\ref{int3}) are
transversal.

\end{appendix}

\end{document}